\documentclass[twocolumn,showpacs,preprintnumbers,amsmath,amssymb]{revtex4}
\usepackage{graphicx}
\begin{document}

%
%
\title{Exploring initial correlations in a Gibbs state by local application of external field}  
\author{Chikako Uchiyama}
\affiliation{%
Faculty of Engineering, University of Yamanashi,
4-3-11 Takeda, Kofu, Yamanashi 400-8511, Japan}
\date{\today} 
%
\def\ch{{\cal H}}
\def\cha{{\cal H}_{S}}
\def\chb{{\cal H}_{E}}
\def\chp{{\cal H}_{P}}
\def\chab{{\cal H}_{SE}}
\def\ee{|1 \rangle \langle 1|}
\def\gg{|0 \rangle \langle 0|}
\def\eg{|1 \rangle \langle 0|}
\def\ges{|0 \rangle \langle 1|}
\def\muv{{\vec \mu}}
\def\ev{{\vec H}}
\def\hbar{\mathchar'26\mkern -9muh}
\def\ih{\frac{i}{\hbar}}
\def\dh{\frac{1}{\hbar}}
\def\v1{\mathbf{1}}
\def\tit{{\tilde t}}
\def\omegac{{\tilde \omega}_{c}}
\def\tbeta{{\tilde \beta}}
\def\am{A_{(m)}(t)}
\def\mm{{\vec \mu}_{(m)}(t)}
\def\thm{\phi_{(m)}(t)}
\def\Xim{\Xi_{(m)}(t)}
\def\xim{\xi_{(m)}(t)}
\def\com{c_{(m)}(t)}
\def\am{A_{(m)}(t)}
\def\amt{A_{(m)}({\kern 1pt}{\tilde t}{\kern 1pt})}
\def\mm{{\vec \mu}_{(m)}(t)}
\def\mmt{\mu_{(m)}({\kern 1pt}{\tilde t}{\kern 1pt})}
\def\thm{\phi_{(m)}(t)}
\def\thmt{\phi_{(m)}({\kern 1pt}{\tilde t}{\kern 1pt})}
\def\Xim{\Xi_{(m)}(t)}
\def\xim{\xi_{(m)}(t)}
\def\nbar{{\bar n}}
\begin{abstract}
We demonstrate that local application of a weak external field increases distinguishability between states with and without initial correlations.  We consider the case where a two-level system linearly and adiabatically interacts with an infinite number of bosons.  We evaluate the trace distance (or, equivalently, the Hilbert-Schmidt distance) between the quantum states which evolve from two kinds of initial states; the correlated Gibbs state and its uncorrelated marginal state.  We find that the trace distance increases above its initial value for any and all parameter settings.  This indicates that we can explore the existence of initial system-environment correlations with applying a local external field, which causes the breakdown of the contractivity.
\end{abstract}
\maketitle
\section{Introduction}
\label{sec:1}
Reduced dynamics of open systems plays an important role in various fields such as condensed matter, quantum optics, and quantum electrodynamics\cite{Breuer}.  
In order to describe the reduced dynamics, an initial condition is often used where the relevant system is statistically independent from its environment.  
The uncorrelated condition was originally used to describe NMR (nuclear magnetic resonance) phenomena for systems with weak system-environment interactions\cite{Wangsness,separable}.  
Besides the initial condition, the Markovian approximation has been used to describe the relaxation phenomena, which gives the so-called Bloch equation\cite{Bloch}.  Completely positive dynamical semigroups have been devised to ascertain decay properties in the Markovian process\cite{Gorini}, imparting contractivity to the open system dynamics.  While dynamical semigroups became the starting point for design of quantum information processing\cite{Nielsen}, as experimental techniques developed to observe non-Markovian dynamics\cite{fwm}, theoretical treatments for such behavior\cite{Breuer} attracted much attention.  The projection operator methods have provided formulations for initially correlated systems\cite{NZ,Mori,SHU}, but the initial correlations have been frequently ignored\cite{nm}.

The effects of initial system-environment correlations have been discussed regarding quantum measurements\cite{Hakim,Smith,Grabert,Karrlein,Ford01,Ford01-2,Lutz,Banerjee,Ankerhold,vanKampen,Bellomo,Ambegaokar,Pollak}, 
complete positivity of dynamical maps\cite{Pechukas,Alicki,Stelmachovic,Jordan,Carteret,Cesar,Shabani}, non-Markovian dynamics\cite{Breuer95,Tan,Ban}, linear-response absorption line shapes \cite{Burnett,Chang,uchiyama09}, 
and general theoretical treatments using the projection operator methods\cite{Royer}.  
Especially relevant has been the finding that initial system-environment correlations require extension of the 
conceptual framework of open system reduced dynamics to include generalization of contractivity \cite{Laine10}.  

The breakdown of contractivity by including initial correlations has been examined by evaluating the time evolution of the trace distance for finite size of environments\cite{Laine10,Smirne10}, which shows an increase over the initial values.  
Dajka, Luczka, and H\"{a}nggi compared four different distance measures: the trace (which is equivalent to Hilbert-Schmidt),  Bures, Hellinger, and quantum Jensen-Shannon distances, and found that only the trace distance reveals the distance increase for a system that initially correlates with an infinite size of environment\cite{Hanggi}.  However, the correlated state in \cite{Hanggi} requires elaborate preparation by quantum engineering.  As a correlated state, we can consider a Gibbs state which includes infinite size of environment.  When we analyze NMR and ESR (electron spin resonance) experiments, or design quantum information processing for condensed matter, it might be necessary to find the system-environment correlations that are inherent to the matter.  A recent study on the transient linear response of a matter system to a suddenly-applied weak external field showed dependence on the initial conditions, correlated Gibbs state, and a conventional factorized state where the relevant system and the environment stay in their equilibrium states\cite{uchiyama09-2}.  Conversely, one might say that the transient linear response of a matter system is useful to detect the initial correlations.  However, in \cite{uchiyama09-2}, it was difficult to distinguish the effect of initial correlations except for strong system-environment interactions and for intermediate temperatures.  In addition, within this approach, we cannot discuss the generalization of contractivity.
 
In this paper, we evaluate the time evolution of trace distance for a matter system where a two-level system interacts with an infinite number of bosons as the environmental system.  We consider that the two-level system linearly and adiabatically interacts with the environment, and that a weak external field is suddenly applied to the two-level system.  We show that the trace distance between the time evolution from the Gibbs state, \(\rho_{SE}\), and from an uncorrelated initial condition clearly exhibits effects of initial correlations as an increase in the short time region. As opposed to \cite{uchiyama09-2}, as an uncorrelated initial condition, we consider the marginal state of the Gibbs state, namely \(\rho_{S} \otimes \rho_{E}\) with \(\rho_{S}={\rm Tr_{E} } \rho_{SE}\) and \(\rho_{E}={\rm Tr_{S} } \rho_{SE}\) where \({\rm Tr_{S}} ({\rm Tr_{E}})\) is a partial trace over the relevant (environmental) system, respectively.  By describing an uncorrelated state as the marginal state, the relevant system interacts with the same environmental features as an average.  We find that the trace distance increases in the short time region even when the system-environment interaction is weak.  This indicates that the local application of a weak external field can cause information inaccessible at an initial time to flow into the relevant system from its infinitely-sized environment included in a Gibbs state, and we can monitor it with the trace distance effectively.

The paper is organized as follows.  In Sec. II, we provide our formulation to obtain the time evolution of the two-level system which linearly and adiabatically interacts with an infinite number of bosons under a suddenly-applied weak external field.  We evaluate the induced dipole moment of the two-level system in Sec. III, and the trace distance time evolution from two different initial conditions in Sec. IV.  We state conclusions in Sec. V. 

\section{Formulation}
\label{sec:2}
We consider a matter system which consists of a two-level system and its environment of an infinite number of bosons.  We assume that the system-environment interaction is linear and adiabatic, which causes the pure dephasing phenomena to the two-level system. 
 The Hamiltonan of the matter system is written as
\begin{equation}
\ch= \cha+\chb+\chab,
\label{eqn:1}
\end{equation} 
with
\begin{eqnarray}
\cha &=& E_{1} \ee +E_{0} \gg, \nonumber \\
\chb &=& \sum_{k} \hbar \omega_{k} b_{k}^{\dagger}b_{k}, \label{eqn:2} \\
\chab &=&  \sum_{k} \hbar g_{k} (b_{k}^{\dagger}+b_{k}) \ee, \nonumber 
\end{eqnarray} 
where \(E_{0}\) and \(E_{1}\) are the energy of the lower and upper states, respectively, of the relevant system, \(\omega_{k}\) is the frequency of the bosonic bath mode, \(b_{k}^{\dagger}\) and \(b_{k}\) are its creation and annihilation operators, and \(g_{k}\) is the coupling strength.  We study the transient behavior of a two-level system after a sudden application of an external field.  The matter-field interaction Hamiltonian is given by
\begin{equation}
\chp(t) = -\frac{1}{2} \muv \cdot \ev \;\theta(t)\eg e^{-i \omega_{p} t}  -\frac{1}{2} \muv^{*} \cdot \ev \; \theta(t) \ges e^{i \omega_{p} t} ,
\label{eqn:3} 
\end{equation} 
where \(\muv\) is the transition dipole moment, \(\ev\) is the amplitude of the external field, \(\omega_{p}\) is the frequency of the external field, and \(\theta(t)\) is the step function. 

In order to evaluate the time evolution of the induced dipole moment, we use a canonical transformation in terms of \(S \equiv \exp[B \ee]\) with \(B \equiv \sum_{k} (g_{k} / \omega_{k}) (b_{k}-b_{k}^{\dagger}) \), since we can eliminate the system-bath interaction from the matter Hamiltonian in the form
\begin{eqnarray}
\ch'&=&S^{\dagger} \ch S = \cha'+\chb, \label{eqn:4} \\
\cha'&=&  E_{1}' \ee + E_{0} \gg,    \label{eqn:5} 
\end{eqnarray} 
with \(E_{1}' \equiv E_{1} -\hbar \sum_{k} (g_{k}^2 /\omega_{k})\). The matter-field interaction is transformed as
\begin{eqnarray}
\chp'(t)&=& S^{\dagger} \chp(t) S \nonumber \\
&=& -\frac{1}{2} \muv \cdot \ev \theta(t) \eg e^{-i \omega_{p} t} G(0) + h. c.,
\label{eqn:6} 
\end{eqnarray} 
with \(G(0) \equiv e^{B^{\dagger}}\).
From Eqs.~(\ref{eqn:4})--~(\ref{eqn:6}), the time evolution of the matter system takes the form
\begin{equation}
\rho(t) =U(t) \rho(0) U^{\dagger} (t)= S e^{-\ih \ch' t} V(t) \rho'(0) V^{\dagger} (t) e^{\ih \ch' t} S^{\dagger},
\label{eqn:7} 
\end{equation} 
where we define the transformed initial condition as \(\rho'(0)\) and 
\begin{eqnarray}
U(t) &\equiv& T_{+} \exp[- (\ih \ch t + \int_{0}^{t} dt' \ih \chp(t'))],  \label{eqn:8}  \\
V(t) &\equiv& T_{+} \exp[-\ih \int_{0}^{t} {\hat {\cal H}}'_{P}(t') dt']
\label{eqn:9} 
\end{eqnarray} 
 with \( {\hat {\cal H}}'_{P}(t) \equiv e^{\ih \ch' t} \chp'(t) e^{-\ih \ch' t}\)\cite{fp}.  In the following, we consider the application of a weak external field, which enables us to approximate, up to first order, the matter-field Hamiltonian in \(V(t)\) as \(V(t) \approx 1-\ih \int_{0}^{t} {\hat {\cal H}}'_{P}(t') dt'\).

\subsection{correlated initial condition - Gibbs state -} 
As a correlated initial condition, we consider that the whole matter system is in a Gibbs state, \(\rho(0)=  \rho_{SE} \equiv \frac{1}{Z} \exp[-\beta \ch] \).
Transforming the correlated initial state, we find \(\rho'(0)= \frac{1}{Z} \exp[-\beta \cha'] \exp[-\beta \chb] \) with \(Z={\rm Tr }[\exp[-\beta \ch]] \) where \({\rm Tr}\) is the trace operation for the total system, which gives 
\begin{equation}
\rho'(0) \equiv S^{\dagger} \rho(0)  S = \rho'_{0} \gg + \rho'_{1} \ee,
\label{eqn:10} 
\end{equation} 
with
\begin{equation}
\rho'_{0} = \frac{e^{-\beta E_{0}} {\tilde \rho}_{E} }{Z_{S}'}, \;\;\;\;
\rho'_{1} = \frac{e^{-\beta E_{1}'} {\tilde \rho}_{E} }{Z_{S}'}.
\label{eqn:11} 
\end{equation} 
In Eq.~(\ref{eqn:11}), we define  \(Z_{S}'={\rm Tr_{S} }[\exp[-\beta \cha']] \) and \({\tilde \rho}_{E} = \exp[-\beta \chb] /Z_{E}\) with \(Z_{E}={\rm Tr_{E}} \exp[- \beta \chb]\).  
Note that the transformation of the correlated state gives a factorized state of the Gibbs state and the bosonic environment, and endows the two-level system with renormalized energies.

The transformed initial condition gives the elements of the reduced statistical operator as
\begin{eqnarray}
{\Bigr [}{\rm Tr_{E}} [U(t) \rho_{SE} U(t)] {\Bigr ]}_{11} &=& \frac{e^{-\beta E_{1}'} }{Z_{S}'}, \nonumber \\
{\Bigr [}{\rm Tr_{E}} [U(t) \rho_{SE} U(t)]{\Bigr ]}_{10} &=& \frac{i \muv \cdot \ev}{2 \hbar}  e^{- i \omega_{p} t} A_{({\rm corr})} (t), \label{eqn:12} 
\end{eqnarray} 
where we define
\begin{eqnarray}
A_{({\rm corr})} (t) &=& \int_{0}^{t} dt'  e^{- i \Delta \omega (t-t')}   \nonumber \\
&&\hspace{-1cm} \times\;( \frac{e^{-\beta E_{0}} }{Z_{S}'}  \Psi_{1} (t-t') - \frac{e^{-\beta E_{1}'} }{Z_{S}'} \Psi^{*}(t-t')  ), \label{eqn:13} 
\end{eqnarray} 
with \(\Delta \omega \equiv ( E_{1}' - E_{0} ) / \hbar - \omega_{p} \).  In  Eq.~(\ref{eqn:13}), we define
\begin{equation}
\Psi_{1} (t-t') \equiv \langle G^{\dagger}(t) G(t')  \rangle = \langle G(t') G^{\dagger}(t) \rangle^{*},
\label{eqn:14} 
\end{equation} 
with \(G(t) \equiv e^{\ih \ch' t} e^{B^{\dagger}} e^{-\ih \ch' t} \), and \(\langle X \rangle \equiv {\rm Tr_{E}} {\tilde \rho}_{E} X \) for an arbitrary operator \(X\).

\subsection{uncorrelated initial condition  - marginal state -}
As an uncorrelated initial state, we take a factorized state, \(\rho(0)= \rho_{S} \otimes \rho_{E}\) which is the marginal of \(\rho_{SE}\) with \(\rho_{S}={\rm Tr_{E} } \rho_{SE}\) and \(\rho_{E}={\rm Tr_{S} } \rho_{SE}\).  (Note the difference between \(\rho_{E}\), and \({\tilde \rho}_{E}\)). As shown in the Appendix, the transformation of the marginal state gives
\begin{equation}
\rho'_{0} = \frac{1}{Z_{S}'} e^{-\beta E_{0}} \rho_{E},\;\;\;\;
\rho'_{1} =  \frac{1}{Z_{S}'} e^{-\beta E_{1}'} G(0) \rho_{E} G^{\dagger}(0), 
\label{eqn:15} 
\end{equation} 
with \( \rho_{E}= \frac{1}{Z_{S}'} (e^{-\beta E_{0}} {\tilde \rho}_{E}  +  e^{-\beta E_{1}'} G^{\dagger}(0) {\tilde \rho}_{E} G(0))\). 
The elements of the reduced statistical operator are obtained as
\begin{eqnarray}
{\Bigr [}{\rm Tr_{E}} [U(t) \rho_{S} \otimes \rho_{E} U(t)] {\Bigr ]}_{11} &=& \frac{e^{-\beta E_{1}'} }{Z_{S}'}, \;\; \nonumber \\
{\Bigr [}{\rm Tr_{E}} [U(t) \rho_{S} \otimes \rho_{E} U(t)] {\Bigr ]}_{10} &=& \frac{i \muv \cdot \ev}{2 \hbar}  e^{- i \omega_{p} t} A_{({\rm marg})} (t), \nonumber \\ \label{eqn:16} 
\end{eqnarray} 
where we define
\begin{eqnarray}
A_{({\rm marg})} (t)&=& \int_{0}^{{\tilde t}} d{\tilde t}'  e^{- i \Delta {\tilde \omega} ({\tilde t}-{\tilde t}')}  \nonumber \\
&&\hspace{-0.5cm} \times \{\frac{e^{-\beta E_{0}}}{Z_{S}'}  ( \frac{e^{-\beta E_{0}}}{Z_{S}'} {\tilde \Psi}_{1} (t-t')  + \frac{e^{-\beta E_{1}'}}{Z_{S}'} {\tilde \Psi}_{2}(t,t')) \nonumber \\
&&-\frac{e^{-\beta E_{1}'} }{Z_{S}'} ( \frac{e^{-\beta E_{0}}}{Z_{S}'}  {\tilde \Psi}_{2}^{*} (t,t') +\frac{e^{-\beta E_{1}'}}{Z_{S}'} {\tilde \Psi}^{*}_{1}(t-t') \},  \nonumber \\
\label{eqn:17} 
\end{eqnarray} 
with
\begin{eqnarray}
\Psi_{2}(t,t') &\equiv& \langle G(0) G^{\dagger}(t) G(t') G^{\dagger}(0) \rangle \nonumber \\
&=& \langle G^{\dagger}(0) G(t') G^{\dagger}(t) G(0) \rangle^{*}.
\label{eqn:18} 
\end{eqnarray}

\section{Time evolution of induced dipole moment}
We evaluate the induced dipole moment under the application of an external field given by
\begin{equation}
\muv(t)={\rm Tr}[(\muv \eg+\muv^{*} \ges) \rho(t)],
\label{eqn:19} 
\end{equation}
using the formulation in the previous section.  Apart from a factor of \(2\vec \mu (\muv^{*} \cdot \ev) / \hbar\), we define the induced dipole moment for the correlated initial condition as \(\mu_{{\rm (corr)}}(t) \) and for the uncorrelated marginal initial condition 
 as \(\mu_{{\rm (marg)}}(t) \).  We obtain
\begin{equation}
\mu_{(m)}(t)  =   |\am| \cos (\omega _p t - \thm ),
\label{eqn:20} 
\end{equation}
where \(A_{{\rm (corr)}}(t) \) and \(A_{{\rm (marg)}}(t) \) are defined in Eqs.~(\ref{eqn:13}) and (\ref{eqn:17}), respectively.  \(\thm\) is the argument of \(\am\). Defining the coupling spectral density as \(h(\omega) \equiv \sum_{k} g_{k}^2 \delta(\omega-\omega_{k})\), we obtain
\begin{eqnarray}
\Psi_{1} (t)&=&\exp[-\int_{0}^{\infty} d\omega \frac{h(\omega)} {\omega^2}  \nonumber \\
                          &&\hspace{0.5cm}   \times \{(1+2 n(\omega)) (1-\cos(\omega t)) + i \sin( \omega t )\}],  \nonumber \\
\label{eqn:21} 
\end{eqnarray}
and
\begin{eqnarray}
\Psi_{2} (t,t') &=& \Psi_{1} (t-t') \nonumber \\
&& \times \exp[-\int_{0}^{\infty} d\omega \frac{h(\omega)} {\omega^2} 2 i (\sin( \omega t')-\sin( \omega t)) ]. \nonumber  \\ \label{eqn:22}
\end{eqnarray} 

Let us now set the spectral density to be Ohmic, i.e., \(h(\omega) \equiv s \omega e^{-\omega / \omega_{c}}\), with the coupling strength, \(s\), and the cut-off frequency, \(\omega_{c}\).  In this case, the renormalized energy is given by \(E_{1}' = E_{1} -\hbar s \omega_{c}\), and we find
\begin{equation}
\Psi_{2} (t,t') = \Psi_{1} (t-t')\exp[- 2 i s ( \arctan \left( {\omega _c \,t'} \right) -\arctan \left( {\omega _c \,t} \right)) ].
\label{eqn:23} 
\end{equation}

\begin{figure}[h]
\begin{center}
\includegraphics[scale=0.45]{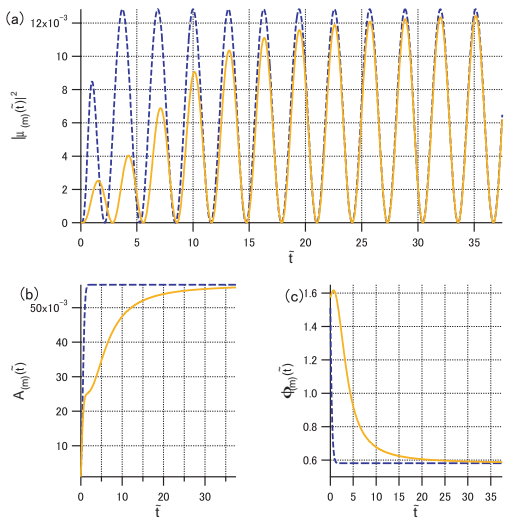}
\end{center}
\caption{(color online) Time evolution of the induced dipole moment for \(k_{B} T=10 \, \hbar \omega_{0}\), \(s=1\), \(\omega_{c} =\omega_{0}/5\), and \(\omega_{p}=\omega_{0}\). Dashed (blue), and light gray solid (orange) lines represent the time evolutions of induced dipole moment for correlated and marginal initial condition, respectively.  Time is scaled as \({\tilde t}=\omega_{0} t\). (a) time evolution of intensity of dipole moment \(|\mmt|^2\), (b) the time evolution of amplitude \(|\amt|\) and (c) the time evolution of phase, \(\thmt\) with \(m={\rm corr\;\;} {\rm or\;\;}  {\rm marg}\).}
\label{fig:fig1}
\end{figure}

In Fig.~\ref{fig:fig1}, we show the time evolution of the intensity of the dipole moment under the application of an external field for \(k_{B} T=10 \,\hbar \omega_{0}\), \(s=1\), \(\omega_{c} =\omega_{0}/5\), and \(\omega_{p}=\omega_{0}\). Time is scaled as \({\tilde t}=\omega_{0} t\). Dashed (blue), and light gray solid (orange) lines represent the time evolutions of induced dipole moment for the correlated and marginal initial conditions, respectively.

We find that the induced dipole moment for the correlated initial condition approaches a stationary oscillation faster than for the marginal initial condition.  As shown in Fig.~\ref{fig:fig1}(b) and (c), the phase and amplitude of the induced dipole moment for the marginal initial condition approaches the same phase and amplitude of the correlated initial condition.  We can explain the long-time behavior using the monotonic time dependence of the \(\arctan \left( {\omega _c \,t} \right) \) function.  The function approaches \(\pi/2\) with increasing time, which means that \(\Psi_{2} (t,t')\) approaches \(\Psi_{1} (t-t')\) for large \(t\).  Comparison between Eqs.~(\ref{eqn:13}) and~(\ref{eqn:17}) reveals that the induced dipole moment for the correlated initial condition agrees with that for marginal initial condition at long times.

\begin{figure}[h]
\begin{center}
\includegraphics[scale=0.45]{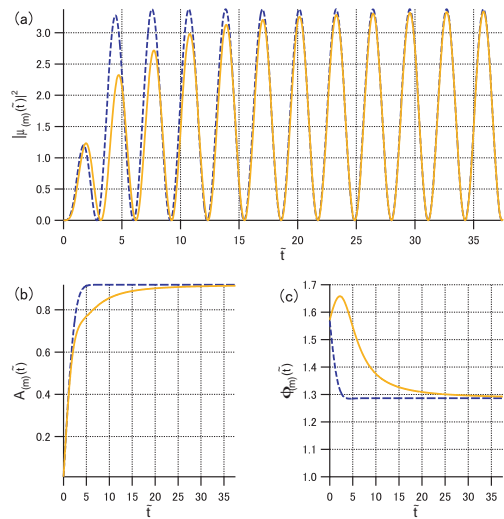}
\end{center}
\caption{(color online) Time evolution of the induced dipole moment for \(k_{B} T=\hbar \omega_{0}\). Other parameters and evaluated quantities are the same as in Fig.~\ref{fig:fig1}.}
\label{fig:fig2}
\end{figure}

\begin{figure}[h]
\begin{center}
\includegraphics[scale=0.45]{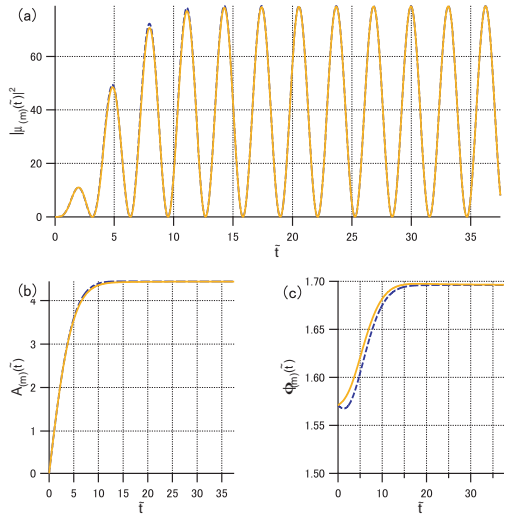}
\end{center}
\caption{(color online) Time evolution of the induced dipole moment for \(k_{B} T=\hbar \omega_{0}/5\). Other parameters and evaluated quantities are the same as in Fig.~\ref{fig:fig1}.}
\label{fig:fig3}
\end{figure}

The time evolutions of the induced dipole moment for lower temperatures of \(k_{B} T=\hbar \omega_{0}\) and \(k_{B} T=\hbar \omega_{0}/5\) are shown in Figs.~\ref{fig:fig2} and~\ref{fig:fig3}, respectively.  As the temperature decreases, we find that the induced dipole moment for the correlated initial condition approaches a stationary oscillation slower, and find a decreased dependence of the time evolution on the initial condition. In the above evaluation, the stationary oscillation for the marginal initial condition coincides with that for the correlated initial condition.  This coincidence is in contrast to the behavior seen with the factorized initial condition, where the system and environment each stay in equilibrium, as discussed in \cite{uchiyama09-2}.

\section{Trace distance}
\label{sec:3}
We found the difference between the time evolution of the induced dipole moment from the Gibbs state of the whole matter system and its marginal state in the previous section.  However, it might be difficult to observe the difference between these time evolutions clearly.  Moreover, we cannot discuss contractivity from the evaluations in the previous section.  

In order to clarify the effects of initial system-environment correlations on open system dynamics, we need a sensitive and tractable measure.  Since the effects of initial correlations appear in time evolutions where non-Markovian features dominate, we can use the trace distance as a measure of non-Markovianity \cite{Breuer09,Laine09}. The trace distance for two quantum states expressed with trace class operators \(\rho^{1}\) and \(\rho^{2}\) is defined as \(D(\rho^{1},\rho^{2})=\frac{1}{2} {\rm Tr} |\rho^{1} - \rho^{2}|\)\cite{Nielsen}.  Here we want to obtain the distance between the reduced dynamics of the two-level system for two kinds of initial conditions: the Gibbs state \(\rho^{1}(0)=\rho_{SE}\), as a correlated initial condition, and its marginal state, \(\rho^{2}(0)=\rho_{S} \otimes \rho_{E}\), as an uncorrelated initial condition.  When the trace distance increases above an initial value, namely, \(D(\rho^{1}(t),\rho^{2}(t)) > D(\rho^{1}(0),\rho^{2}(0))\), we deduce that information, which is inaccessible at the initial time, flows into the relevant system through system-environment correlations.  Moreover, this increase indicates a breakdown of the contractivity\cite{Laine10}.

For our model, we obtain the trace distance in the form
\begin{eqnarray}
D(Tr_{E}[ U(t) \rho_{SE} U^{\dagger}(t)], Tr_{E}[ U(t) \rho_{S} \otimes \rho_{E} U^{\dagger}(t)]) \nonumber \\ 
&&\hspace{-8cm} = \sqrt{ |{\cal M}_{11} |^2 + |{\cal M}_{10}|^2},\label{eqn:24} 
\end{eqnarray} 
with \({\cal M}={\rm Tr_{E}} [U(t) \rho_{SE} U^{\dagger}(t)]- {\rm Tr_{E}} [U(t) \rho_{S} \otimes \rho_{E} U^{\dagger}(t)] \).

Using Eqs.~(\ref{eqn:12}) and Eq.~(\ref{eqn:16}), we obtain the trace distance as
\begin{eqnarray}
D(Tr_{E}[ U_{t} \rho_{SE} U_{t}^{\dagger}], Tr_{E}[ U_{t} \rho_{S} \otimes \rho_{E} U_{t}^{\dagger}]) = |{\cal M}_{10} |.
\label{eqn:25} 
\end{eqnarray} 

\begin{figure}[h]
\begin{center}
\includegraphics[scale=0.4]{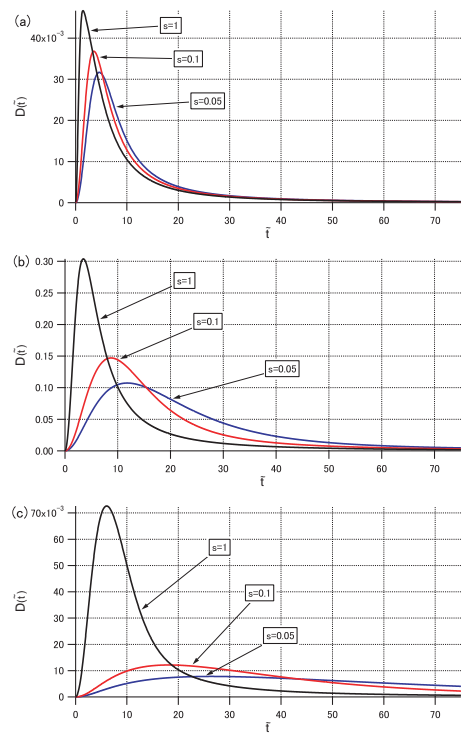}
\end{center}
\caption{(color online) Time evolution of the trace distance for \(\omega_{c} =\omega_{0}/5\), and \(\omega_{p}=\omega_{0}\) with changing temperature as (a)\(k_{B} T=10 \, \hbar \omega_{0}\), (b)\(k_{B} T=\, \hbar \omega_{0}\), and (c)\(k_{B} T=\hbar \omega_{0} / 5\). In each figure, black, light gray (red), and dark gray (blue) lines correspond to the case of \(s=1\), \(0.1\), and \(0.05\), respectively.}
\label{fig:fig4}
\end{figure}

We show the time evolution of the trace distance, apart from the dimensionless quantity \(\frac{\muv \cdot \ev}{2 \hbar \omega_{0}}\), for Ohmic spectral density in Fig.~\ref{fig:fig4} by setting \(\omega_{c} =\omega_{0}/5\), and \(\omega_{p}=\omega_{0}\).  Here we scaled time variable as \({\tilde t} \; ( \equiv \omega_{0} t)\).

Figures~\ref{fig:fig4}(a)--(c) correspond to the trace distance time evolution at various temperatures: (a)\(k_{B} T=10 \, \hbar \omega_{0}\), (b)\(k_{B} T=\, \hbar \omega_{0}\), and (c)\(k_{B} T=\hbar \omega_{0} / 5\).  In each figure, we vary \(s\) as \(1\), \(0.1\), and \(0.05\).  We find in Fig.~\ref{fig:fig4}(a) that the trace distance increases in the short time region, which signifies the breakdown of contractivity for each value of \(s\).  In Fig.~\ref{fig:fig4}(a), we also find the trace distances approach zero at long times, which arises from the fact that the \(\arctan \left( {\omega _c \,t} \right) \) function in Eq.~(\ref{eqn:23}) monotonically approaches \(\pi/2\) with increasing time and \(\Psi_{2} (t,t')\) approaches \(\Psi_{1} (t-t')\) for large \(t\).  For lower temperature cases as shown in Figs.~\ref{fig:fig4}(b) and (c), we find qualitatively similar behavior as in Fig.~\ref{fig:fig4}(a). Comparing these figures, we find that the peak value of the trace distance becomes larger for increasing values of \(s\) and at intermediate temperatures.  

In the above evaluations, we find that the reduced dynamics of the matter system under a weak external field shows a breakdown of the contractivity for any parameter setting, which allows us a method to monitor the initial system-environment correlations in a Gibbs state.
 
\section{Conclusions}
\label{sec:4}
We have studied the time evolution of a matter system which is composed of a two-level system and a bosonic environment of infinite size, under a suddenly-applied weak external field.  
Assuming the system-environment interaction to be linear and adiabatic, we evaluated the induced dipole moment for two kinds of initial conditions: a correlated Gibbs state and an uncorrelated marginal state.  
We found that the time evolution of the induced dipole moment depends on the initial condition.  However, the difference depends on the parameter settings and, under certain settings, may be difficult to observe.  
By evaluating the trace distance in the reduced dynamics for these two initial conditions, we found that we can overcome the parameter dependence.  
The trace distance clearly increases in the short time region for cases of arbitrary system-environment interactions and temperatures.  
We can effectively monitor the initial system-environment correlations in a Gibbs state with the breakdown of the contractivity.

\begin{acknowledgements}
In memory of former Prof. Hajime Mori.
\end{acknowledgements}

\appendix *
\section{Derivation of Eq.~(\ref{eqn:15})}
In this appendix, we explain how to obtain the marginal initial state 
\begin{equation}
\rho(0)= \rho_{S} \otimes \rho_{E}\
\label{eqn:A1} 
\end{equation} 
with \(\rho_{S}={\rm Tr_{E} } \rho_{SE}\) and \(\rho_{E}={\rm Tr_{S} } \rho_{SE}\) for \(\rho_{SE} (= \frac{1}{Z} \exp[-\beta \ch]) \). 
The correlated state \(\rho_{SE}\) is rewritten as
\begin{eqnarray}
\rho_{SE} \nonumber \\
&&\hspace{-1cm} = \frac{1}{Z} \exp[-\beta \ch_{0}] T_{+} \exp[- \int_{0}^{\beta} d \lambda e^{\lambda \ch_{0}} \chab e^{-\lambda \ch_{0}}] \nonumber \\
&&\hspace{-1cm} =  \frac{1}{Z} ( e^{-\beta E_{0}} \exp[-\beta \ch_{E}]   \gg \nonumber\\
&&\hspace{-0.5cm} + e^{-\beta E_{1}} \exp[-\beta \ch_{E}] \nonumber \\
&&\hspace{-0.5cm} \times T_{+} \exp[-\sum_{k}  \hbar g_{k} \int_{0}^{\beta} d \lambda (b_{k}^{\dagger} e^{\hbar \lambda \omega_{k}}+ b_{k} e^{-\hbar \lambda \omega_{k}})] \ee), \nonumber \\
\label{eqn:A2} 
\end{eqnarray} 
where we define \(\ch_{0} \equiv \cha + \chb \) and use the \(T_{+}\) operator to order \(\lambda\) from right to left.  Using the relation as 
\begin{eqnarray}
\exp[-\beta \ch_{E}] \; T_{+} \exp[-\sum_{k} \hbar g_{k} \int_{0}^{\beta} d \lambda (b_{k}^{\dagger} e^{\hbar \lambda \omega_{k}}+ b_{k} e^{-\hbar \lambda \omega_{k}}) ]  \nonumber \\
&&\hspace{-9cm} = \exp[-\beta (\ch_{E}+\sum_{k} \hbar g_{k} (b_{k}^{\dagger}+b_{k}))]  \nonumber \\
&&\hspace{-9cm} = G^{\dagger}(0) \exp[-\beta (\ch_{E} - \hbar \sum_{k} \frac{g_{k}^2}{\omega_{k}})] G(0), 
\label{eqn:A3} 
\end{eqnarray}
we obtain
\begin{equation}
\rho_{SE} = \frac{1}{Z_{S}'} (e^{-\beta E_{0}} {\tilde \rho}_{E} \gg+ e^{-\beta E_{1}'} G^{\dagger}(0) {\tilde \rho}_{E} G(0)\ee),
\label{eqn:A4} 
\end{equation} 
which gives the partial trace operation on \(\rho_{SE}\) over the system and the environment as
\begin{eqnarray}
\rho_{S}&=& \frac{1}{Z_{S}'} (e^{-\beta E_{0}} \gg+ e^{-\beta E_{1}'}  \ee), \label{eqn:A5} \\
\rho_{E}&=&\frac{1}{Z_{S}'}  ( e^{-\beta E_{0}} {\tilde \rho}_{E}  + e^{-\beta E_{1}'} G^{\dagger}(0) {\tilde \rho}_{E} G(0)).
\label{eqn:A6} 
\end{eqnarray} 
From these, we obtain the marginal state as
\begin{equation}
\rho_{S} \otimes \rho_{E} = \frac{1}{Z_{S}'} (e^{-\beta E_{0}} \rho_{E} \gg + e^{-\beta E_{1}'} \rho_{E} \ee),
\label{eqn:A7} 
\end{equation} 
which is transformed into
\begin{eqnarray}
( \rho_{S} \otimes \rho_{E} )' &=& \rho'_{0} \gg + \rho'_{1} \ee \nonumber \\
&=& \frac{1}{Z_{S}'} (e^{-\beta E_{0}} \rho_{E}\gg \nonumber \\
&& + e^{-\beta E_{1}'}  G(0) \rho_{E} G^{\dagger}(0) \ee).
\label{eqn:A8} 
\end{eqnarray} 
Equation~(\ref{eqn:A8}) gives Eq.~(\ref{eqn:15}).



\begin{thebibliography}{99} 
\bibitem{Breuer} H. P. Breuer, and F. Petruccione, {\it The Theory of Open Quantum Systems} (Oxford University Press, Oxford, 2002).
\bibitem{Wangsness} R. K. Wangsness and F. Bloch, Phys. Rev. {\bf 89}, 728 (1953).
\bibitem{separable}A. Abragam, {\it Principles of Nuclear Magnetism}, (London,Oxford press,1961);C. P. Slichter, {\it Principles of Magnetic Resonance} (Berlin,Springer,1981) and cited therin.
\bibitem{Bloch} F. Bloch, Phys. Rev. {\bf 70}, 460 (1946).
\bibitem{Gorini} V. Gorini, A. Kossakowski and E. C. G. Sudarshan, J. Math. Phys. {\bf 17}, 821 (1976).
\bibitem{Nielsen} M. A. Nielsen and I. L. Chuang, {\it Quantum Computation and Quantum Information} (Cambridge University Press, Cambridge, 2000).
\bibitem{fwm} S. Saikan, J. W.-I. Lin and H. Nemoto, Phys. Rev. B, {\bf 46}, 11125(1992); E. T. J. Nibbering, D. A. Wiersma, K. Duppen, Phys. Rev. Lett., {\bf 66}, 2464 (1991); U. Woggon, F. Gindele, W. Langbein and J. M. Hvam, Phys. Rev. B, {\bf 61}, 1935(2000); Y. Masumoto et. al., Physica E, {\bf 26}, 413(2005).
\bibitem{NZ} S. Nakajima, Prog. Theor. Phys., {\bf 20}, 1338 (1958); R. Zwanzig, J. Chem. Phys.,  {\bf 33}, 1338 (1960).
\bibitem{Mori} H. Mori, Prog. Theor. Phys., {\bf 33} 423 (1965).
\bibitem{SHU} N. Hashitsume, F. Shibata, and M. Shingu, J. Stat. Phys., {\bf 17}
155 (1977); S. Chaturvedi and F. Shibata, Z. Phys. B {\bf 35} 297 (1979); F. Shibata and T. Arimitsu, J. Phys. Soc. Jpn., {\bf 49} 891 (1980); C. Uchiyama and F. Shibata, Phys. Rev. E. {\bf 60}, 2636 (1999).
\bibitem{nm} J. Piilo, S. Maniscalco, K. Harkonen and K. A. Suominen, Phys. Rev. Lett., {\bf 100}, 180402 (2008); J. Piilo, K. Harkonen, S. Maniscalco and K. A. Suominen, Phys. Rev. A {\bf 79}, 062112 (2009); H. -P. Breuer and J. Piilo, Europhys. Lett.,  {\bf 85}, 50004 (2009); H. -P. Breuer and B. Vacchini, Phys. Rev. Lett., {\bf 101}, 140402 (2008); H. -P. Breuer and B. Vacchini, Phys. Rev. E, {\bf 79}, 041147 (2009); H. -P. Breuer, Phys. Rev. A {\bf 75}, 022103 (2007); S. Daffer, K. W\'{o}dkiewicz, J. D. Cresser, and J. K. McIver, Phys. Rev. A {\bf 70}, 010304 (2004); A. Kossakowski and R. Rebolledo, Open Syst. Inf. Dyn., {\bf 15}, 135 (2008); C. Uchiyama and F. Shibata, Phys. Lett. A {\bf 267}, 7 (2000); C. Uchiyama and F. Shibata, J. Phys. Soc. Jpn. {\bf 69}, 2829 (2000); D. Chru\'{s}ci\'{n}ski, A. Kossakowski and \'{A}. Rivas, Phys. Rev. A, {\bf 83}, 052128 (2011).
\bibitem{Hakim} V. Hakim, and V. Ambegaokar, Phys. Rev. A {\bf 32}, 423 (1985).
\bibitem{Smith} C. M. Smith and A. O. Caldeira, Phys. Rev. A {\bf 36}, 3509 (1987).
\bibitem{Grabert} H. Grabert, P. Schramm and G. L. Ingold, Phys. Rep. {\bf 168}, 115 (1988).
\bibitem{Karrlein}R. Karrlein, and H. Grabert, Phys. Rev. E {\bf 55}, 153 (1997).
\bibitem{Ford01}G. W. Ford, J. T. Lewis, and R. F. O'Connell, Phys. Rev. A {\bf 64}, 032101 (2001).
\bibitem{Ford01-2}G. W. Ford, and R. F. O'Connell, Phys. Lett. A {\bf 2865}, 87 (2001); Phys. Rev. D {\bf 64}, 105020 (2001).
\bibitem{Lutz} E. Lutz, Phys. Rev. A {\bf 67}, 022109 (2003).
\bibitem{Banerjee} S. Banerjee and R. Ghosh, Phys. Rev. A {\bf 62}, 042105 (2000); Phys. Rev. E {\bf 67}, 056120 (2003).
\bibitem{Ankerhold} J. Ankerhold, Europhys. Lett. {\bf 61}, 301 (2003).
\bibitem{vanKampen} N. G. van Kampen, J. Stat. Phys. {\bf 115}, 1057 (2004).
\bibitem{Bellomo} B. Bellomo, G. Compagno and F. Petruccione, J. Phys. A {\bf 38},10203 (2005).
\bibitem{Ambegaokar} V. Ambegaokar, J. Stat. Phys. {\bf 125}, 1187 (2006); Ann. Phys. {\bf 16}, 319 (2007).
\bibitem{Pollak}E. Pollak, J. Shao, and D. H. Zhang, Phys. Rev. E {\bf 77}, 021107 (2008).
\bibitem{Pechukas}P. Pechukas, Phys. Rev. Lett. {\bf 73}, 1060 (1994); Phys. Rev. Lett. 75, 3021 (1995).
\bibitem{Alicki}R. Alicki, Phys. Rev. Lett. {\bf 75}, 3020 (1995). 
\bibitem{Stelmachovic}P. Stelmachovic and V. Buzek, Phys. Rev. A {\bf 64}, 062106 (2001);  Phys. Rev. A {\bf 67}, 029902(E) (2003).
\bibitem{Jordan}T. F. Jordan, A. Shaji, and E. C. G. Sudarshan, Phys. Rev. A {\bf 70}, 052110 (2004); A. Shaji and E. C. G. Sudarshan, Phys. Lett. A  {\bf 341}, 48 (2005).
\bibitem{Carteret}H. A. Carteret, D. R. Terno, and K. \.{Z}yczkowski, Phys. Rev. A {\bf 77},  042113 (2008). \bibitem{Cesar} C. A. Rodr\'{i}guez-Rosario, et. al., J. Phys. A  {\bf 41} , 205301 (2008).
\bibitem{Shabani} A. Shabani and D. A. Lidar, Phys. Rev. Lett. {\bf 102}, 100402 (2009).
\bibitem{Breuer95} H. P. Breuer, and F. Petruccione, Phys. Rev. Lett. {\bf 74}, 3788 (1995).
\bibitem{Tan} H.-T. Tan and W.-M. Zhang, Phys. Rev. A {\bf 83},  032102 (2011). 
\bibitem{Ban} M. Ban, S. Kitajima, and F. Shibata, Phys. Lett. A {\bf 375},2283 (2011).
\bibitem{Burnett} P. Thomann, K. Burnett, and J. Cooper, Phys. Rev. Lett. {\bf 45}, 1325 (1980);K. Burnett, J. Cooper, R. J. Ballagh and E. W. Smith, Phys. Rev. A {\bf 22}, 2005 (1980); K. Burnett and J. Cooper, Phys. Rev. A {\bf 22}, 2027 (1980); Phys. Rev. A {\bf 22}, 2044 (1980).
\bibitem{Chang} T.-M. Chang and J. L. Skinner, Physica A {\bf 193}, 483 (1993).
\bibitem{uchiyama09} C. Uchiyama, M. Aihara, M. Saeki, S.Miyashita, Phys. Rev. E, {\bf 80},  021128 (2009); C. Uchiyama, Prog. Theor. Phys. Suppl. {\bf 184},  476 (2010).
\bibitem{Royer}A. Royer, Phys. Rev. Lett. {\bf 77}, 3272 (1996); A. Royer, Phys. Lett. A {\bf 315}, 335 (2003).
\bibitem{Laine10}E. -M. Laine, J.Piilo, and H. -P. Breuer, Europhys. Lett.,  {\bf 92}, 60010 (2010).
\bibitem{Smirne10}A. Smirne, H. -P. Breuer, J. Piilo and B. Vacchini, Phys. Rev. {\bf A}, 82 062114 (2010).
\bibitem{Hanggi} J. Dajka, J. Luczka and P. H\"{a}nggi, Phys. Rev. A {\bf 84}, 032120 (2011) .
\bibitem{uchiyama09-2}C. Uchiyama and M. Aihara, Phys. Rev. A, {\bf 82}, 044104 (2010).
\bibitem{fp} Please note the typos of the signs of time evolution operators in Eq.(11) in \cite{uchiyama09-2}.
\bibitem{Breuer09}H. -P. Breuer, E. -M. Laine, and J.Piilo, Phys. Rev. Lett., {\bf 103}, 210401 (2009).
\bibitem{Laine09}E. -M. Laine, J. Piilo, and H. -P. Breuer, Phys. Rev. A, {\bf 81}, 062115 (2010).
\end{thebibliography}
 \end{document}